\documentclass{optica-article}

\journal{opticajournal} % for journals or Optica Open

\articletype{Research Article}

\usepackage{lineno}
\begin{document}

\title{First x-rays from a compact and tunable LINAC-based Compton scattering source}

\author{I.J.M. van Elk,\authormark{1,*} C.W. Sweers,\authormark{1} D.F.J. Nijhof,\authormark{1} R.G.W. van den Berg,\authormark{1} T.G. Lucas,\authormark{1} X.F.D. Stragier,\authormark{1} P.Tack,\authormark{2} M.N. Boone,\authormark{2} O.J. Luiten,\authormark{1} and  P.H.A. Mutsaers\authormark{1}}

\address{\authormark{1} Department of Applied Physics and Science Education, Coherence and Quantum Technology Group, Eindhoven University of Technology, P.O. Box 513, 5600 MB Eindhoven, The Netherlands\\
\authormark{2} Radiation Physics Research Group - UGCT, Department of Physics and Astronomy, Ghent University, Proeftuinstraat 86/N12, Ghent B-9000, Belgium\\
}
\email{\authormark{*}i.j.m.v.elk@tue.nl} 

\begin{abstract*} 
In this paper, we present the first measurements of x-rays produced with a compact, narrowband, and tunable inverse Compton scattering-based x-ray source, developed at Eindhoven University of Technology. A flux of $1.2 \cdot 10^3$ photons per shot was measured, in agreement with simulations. Using a high-resolution spectral camera, we show that the photon energy can be tuned continuously from 5.8~keV to 10.7~keV with a bandwidth of 4\%. The measured x-ray pulse length was in the picosecond range. Additionally, we show that the source allows full control over the x-ray polarization control. By optimizing experimental parameters, implementing improvements to the setup and further conditioning of the accelerator structure, a brilliance of $10^{12}$ photons/(s $\times$ mrad$^2$ $\times$ mm${^{2}}$ $\times$ 0.1\% BW) can be achieved, with photon energies up to 40 keV. Because the complete electron beamline fits on a single optical table, it is suitable as an in-house x-ray source for university laboratories, industrial production lines, museums, and hospitals.
\end{abstract*}

\vspace{-0.1cm}

\section{Introduction}
Using x-rays to resolve the structure of matter has been essential for science and society for decades. The discovery of x-rays by Röntgen in 1895 was rapidly followed by their application in many research areas, indicating the importance of high-quality x-ray sources \cite{Röntgen_1896,Clark_1929, Behling_2020}. Currently, x-rays sources are still an indispensable tool for medical diagnostics \cite{Berthe_2024,Gradl_2018, Donato_2024} and biological research \cite{Stagno_2017, vanderSchot_2015}. In addition, forensics and security \cite{Meaney_2008,Akcay_2022}, industry \cite{Holler_2017, Bakavos_2010,Sullivan_2015}, material science \cite{Das_2019, Mayo_2012}, and cultural heritage research \cite{Alfeld_2017} greatly benefit from the implementation of x-ray technology.\\
\indent The most common x-ray source is the x-ray tube, which has already been put into practice for over a century \cite{Nascimento_2014}. The working principle of the tube is by targeting an anode metal with accelerated electrons. Over time, developments such as the use of a rotating or liquid metal anode, increased the performance of the device \cite{Behling_2020,Hemberg_2003}. The brilliance, a figure of merit for the output quality, of contemporary tubes is on the order of $10^9-10^{10}$ photons/(s $\times$ mrad$^2$ $\times$ mm${^{2}}$ $\times$ 0.1\% BW) \cite{Coleman_2023,Wansleben_2019}. Unfortunately, this is not sufficient for all applications. The demand for even higher brilliance, meaning x-ray beams with increased photon flux, a smaller divergence, and a narrower bandwidth gave rise to new types of x-ray sources: synchrotrons and x-ray free electron lasers (XFELs).\\
\indent By applying bending magnets, wigglers, and undulators, the accelerator-based facilities reach up to a brilliance of 10$^{25}$ photons/(s $\times$ mrad$^2$ $\times$ mm${^{2}}$ $\times$ 0.1\% BW) \cite{Altarelli_2015}. Moreover, the continuous tunability of the photon energy and ultrashort x-ray pulses featured in these sources opens up new research possibilities, such as K-edge subtraction imaging and pump-probe experiments \cite{Seddon_2017, Hutchison_2024,Levantino_2015}. However, to achieve this these facilities are required to be orders of magnitude larger in size than the most advanced x-ray tube, leading to high construction and maintenance costs. As a result, a combined number of less than fifty synchrotrons and approximately ten XFELs are operational worldwide, limiting the availability for a large group of users \cite{Sorensen_2006,Huang_2021}.\\
\indent To bridge the gap between demand and accessibility of high brilliance beams, recent advancements in accelerator and laser technologies are exploited to develop compact hard x-ray sources. By replacing a standard undulator with a high-power intense laser pulse, inducing the process of inverse Compton scattering (ICS), the size of a synchrotron-like source can be decreased to laboratory scale. Although the large facilities outperform ICS-based sources in terms of x-ray beam quality, the possibility for an in-house source would remove practical burdens for research institutes, companies, and hospitals. Furthermore, unlike conventional x-ray tubes, ICS-based sources are also continuously tunable, generate short x-ray pulses, and offer straighforward polarization control \cite{Zhang_2024}. For this reason, multiple ICS-based x-ray sources are either being developed or in commissioning \cite{Melcher_2024, Dupraz_2020,Gao_2024, Graves_2014 ,Gunther_2020}. \\
\indent Among these ICS projects is Smart*Light, the compact source developed at the Eindhoven University of Technology. Smart*Light incorporates a modified version of an X-band accelerating structure of the Compact Linear Collider (CLIC) project, allowing the complete electron beamline to fit on a three-meter-long optical table. Furthermore, the accelerating structure is adapted to capture 100 keV electron bunches. This permits the use of different types of injectors \cite{vanOudheusden_2007, Toonen_2019, Nijhof_2023}. The modularity enables the user to tailor the output, which can be matched to the needs of the desired application. Smart*Light potentially improves upon x-ray tubes by an order of magnitude in brilliance and by featuring continuous energy tunability from 5 to 40 keV, while restricting the setup size to a single table allows it to still function as an in-house instrument.\\
\indent In this paper we present the first x-ray measurements of Smart*Light, demonstrating the viability of the compact ICS source. The paper is organized as follows: First, we describe the physical principles behind ICS in Section \ref{sec:ICS}. In Section \ref{sec:philosophy}, we address the philosophy behind the Smart*Light project, expressing what makes Smart*Light unique compared to other ICS sources, and why specific choices were made. In Section \ref{sec:SL beamline}, a brief overview of the beamline is given. A more detailed description of the complete beamline will be published in a separate publication. Finally, Section \ref{sec:results} presents the x-ray results, showing the continuous tunability of the photon energies, the characteristic short and narrow band pulses, and the control in polarization.

\section{Principle of ICS}\label{sec:ICS}
\begin{figure}[ht]
\centering\includegraphics[width=0.75\linewidth]{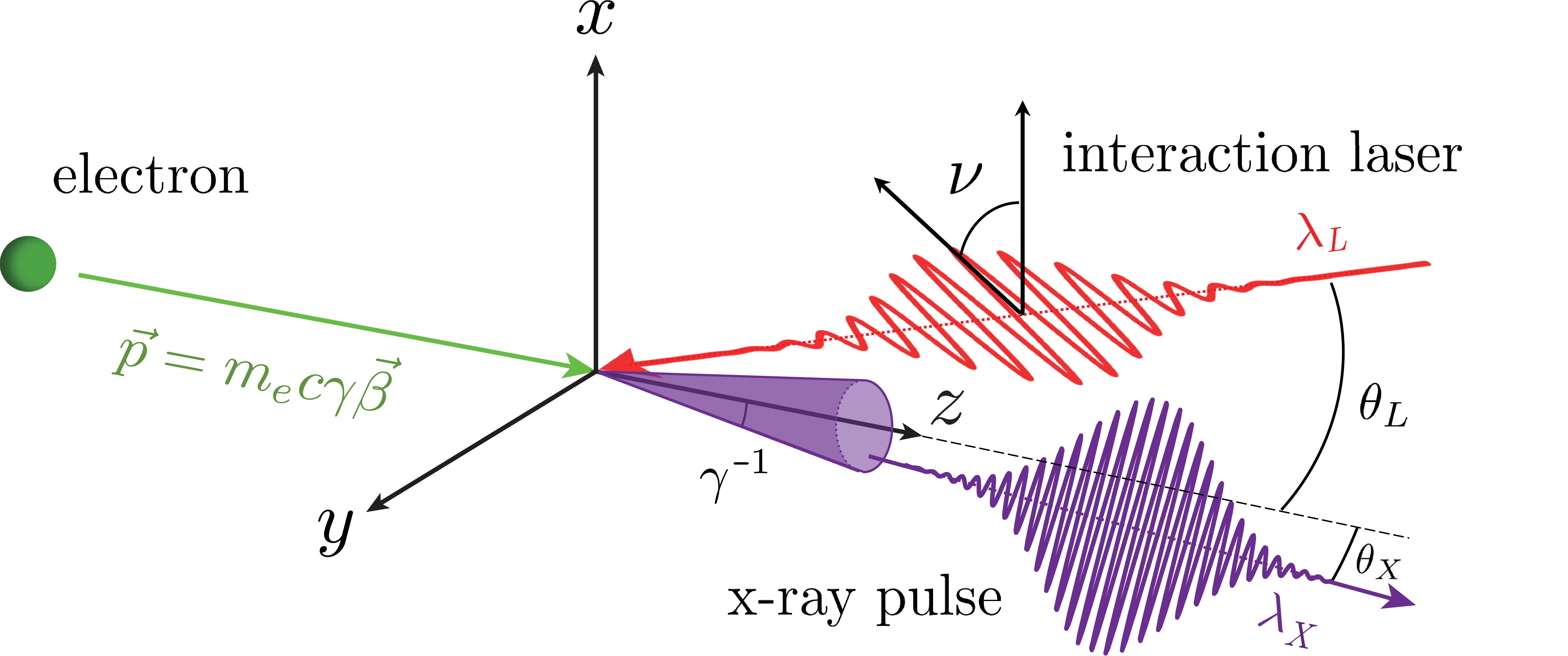}
\caption{Schematic representation of a relativistic electron and a linearly polarized laser pulse inducing ICS, resulting in a light pulse with wavelength $\lambda_X$. The coordinate system has been defined such that the interaction laser propagates in the $y$-$z$ plane, while the electron electron propagates along the $z$-axis.}\label{fig:ICSschematic}
\vspace{-0.3cm}
\end{figure}

In essence, ICS is a process where photons gain energy by scattering off relativistic electrons. This up-conversion is manifested through a double Doppler-shift. A configuration to induce ICS is schematically represented in Fig. \ref{fig:ICSschematic}, where an electron with mass $m_e$, normalized velocity $\vec{\beta}=\vec{v}/c$ with corresponding Lorentz factor $\gamma$ and momentum $\vec{p}=m_ec\gamma \vec{\beta}$ is incident on a high-intensity laser pulse with wavelength $\lambda_L$ and angular frequency $\omega_L$, with $c$ the velocity of light. In the rest frame of the electron, it experiences a laser pulse with an up-shifted frequency. While in the laser pulse, the electron oscillates at this Doppler shifted frequency, upon which it emits dipole radiation. When transformed back to the lab frame, the additional Doppler-shift increases the frequency of the dipole radiation such that a light pulse with wavelength $\lambda_\mathrm{X}$ and angular frequency $\omega_X  = 2\pi c/\lambda_X$ is observed. Assuming a collision angle $\theta_L$ between the laser and a relativistic beam ($\gamma \gg 1$) while neglecting electron recoil, it can be shown that
\begin{equation}\label{eq:wavelength}
    \omega_X = \omega_L \frac{1+\beta \cos \theta_L}{1-\beta \cos \theta_X},
\end{equation}

\noindent where $\theta_X$ is the angle between x-ray emission and the propagation direction of the electron \cite{Ride_1995}. Eq. \eqref{eq:wavelength} only considers linear interactions, i.e., the dimensionless vector potential of the interaction laser $a_0=e E_0/(2 \pi m_e c^2)\ll1$, which holds for the experimental parameters relevant throughout this work. 
Assuming head-on scattering ($\theta_L=0$) and Gaussian laser and electron beam profiles, a simple analytic expression for the x-ray yield $N_X$ can be derived. For the case when the root-mean-square (rms) pulse length of the electron beam $\tau_e$ is smaller than its depth of focus, the electron beam equivalent "Rayleigh length", and the rms pulse length of the laser $\tau_L$ is significantly shorter than its Rayleigh length $z_R$, i.e., $c\tau_L\ll z_R $, $N_X$ is given by:

\begin{equation}\label{eq:Nphotons}
    N_X = \frac{\sigma_T N_e N_L}{2 \pi (\sigma_e^2+\sigma_L^2)}.
\end{equation}

\noindent In Eq. \eqref{eq:Nphotons}, $\sigma_T=8\pi r_0^2/3$ is the Thomson cross-section with $r_0 \approx 2.82$ fm being the classical electron radius, $N_e$ is the number of electrons in the electron pulse, $N_L$ is the number of photons in the laser pulse, $\sigma_e$ the rms-value of the electron beam's transverse size and $\sigma_L$ the rms laser spot size where $\sigma_L=w_0/2$ with $w_0=\sqrt{\lambda_xz_R/\pi}$ the beam waist. As a result of the Doppler shifts, the x-ray emission will be highly directional along the propagation direction of the electrons. It can be shown that for the depicted configuration in Fig. \ref{fig:ICSschematic} with $\theta_L=0$ and for small angles of $\theta_X$, the scattered angular intensity distribution satisfies 

\begin{equation}\label{eq:polchange}
    \frac{dN}{d\Omega dt} \propto \frac{1}{(1+\gamma^2\theta_X^2)^2}\left[ 1-\frac{4\gamma^2\theta_X^2}{(1+\gamma^2\theta_X^2)^2} \cos^2(\nu-\phi) \right]
\end{equation}
with $\nu$ the angle between the polarization and the $x$-axis, and $\phi$ is the angle between the $x$-axis and the direction of the considered emission projected on the $x$-$y$ plane \cite{Brown_2004}. The right-hand component shows that in the plane perpendicular to the polarization, the intensity FWHM lies at $\theta_X =1/\gamma$, while in the plane parallel to the polarization the intensity goes to zero at $\theta_X =1/\gamma$.
\\
\indent To compare the quality of the x-ray source, the brilliance is often used as figure of merit, which is defined as:

\begin{equation}\label{eq:brilliance1}
    B_X = \frac{f N_X}{\Delta A \Delta\Omega\Delta \omega_X/\omega_X}.
\end{equation}

\noindent It describes the amount of x-ray photons $f N_X$ that are generated per second, per unit area $\Delta A$, per unit solid angle $\Delta \Omega$, and per relative energy spread $ \Delta \omega_X/\omega_X$. $f$ denotes the repetition frequency of the x-ray source. It is conventional to use FWHM values, indicated with $\Delta$.

\section{The philosophy behind Smart*Light} \label{sec:philosophy}

The Smart*Light project has been set up with an explicit goal in mind: the development of a compact, tunable, high-brilliance x-ray source applicable for experiments and diagnostics conducted in companies, museums, hospitals, and research institutes. From this approach, the three key elements of Smart*Light's philosophy are derived, forming the guideline upon which decisions were made in the design process. Overall, we opted for a linear accelerator (LINAC) based design to accommodate these key elements, which are detailed below:

\begin{enumerate}
    \item \textbf{Compactness:} the electron beamline should fit on a single 3 m-sized optical table.\\ A table-top setup enables the possibility for an in-house x-ray source, removing schedule and travel barriers. To construct a beamline of this size, it was specifically chosen to use an X-band (8-12 GHz) RF linear accelerator structure over the more conventional S-band (2-4 GHz) and C-band (4-8 GHz) structures. In comparison to the latter two RF-bands, X-band can reach the highest accelerating gradients and higher repetition rates allowing for a more compact design.
   
    \item \textbf{Robustness:} the system should be robust and reliable.\\
    For users it is of importance that the setup operates on demand without the risk that equipment fails or breaks. For this reason, it was chosen to not push the capability of the system to its limits. An example hereof is the generated acceleration gradients in the X-band structure, which could be trained over 100 MV/m. By staying well below this value, dark current is limited and the chance of detrimental vacuum breakdowns is reduced. Moreover, the user can implement a DC photogun as a robust and reliable electron source.
    \item \textbf{Modularity:} elements of the beamline can be added or replaced.\\
    With a modular design, the output of the setup can be tailored to the needs of the desired application. It is for example possible to implement additional accelerator structures to increase the generated x-ray energy. Furthermore, the structure of the currently implemented LINAC was designed to permit injection with 100 keV electron bunches, allowing the user to choose different types of electron sources, rather than being restricted to a single RF photoinjector \cite{Toonen_2019,Nijhof_2023,Pasmans_2016}. 
    
\end{enumerate}

 Each of the elements aims to improve the applicability for the user, emphasizing that Smart*Light is developed to be an in-house instrument, similar to the electron microscope. Users will benefit from Smart*Light's capabilities which combine a synchrotron's tunability with a x-ray tube's accessibility.

\section{Smart*Light setup}\label{sec:SL beamline}

\begin{figure}[htbp]
\centering\includegraphics[width=0.8 \linewidth]{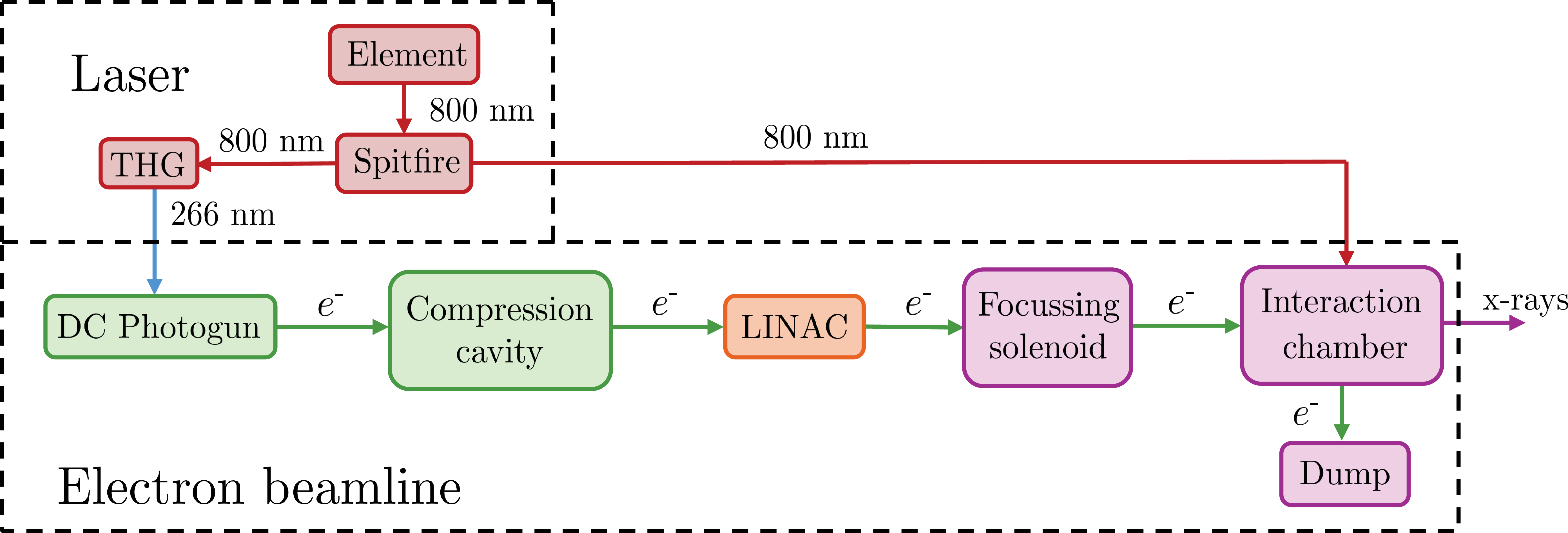}
\caption{Schematic overview of the system, indicating the different segments and related in- and outputs of the Smart*Light setup.}\label{fig:systemoverview}
\end{figure}

\begin{figure}[b]
\vspace{-0.4cm}
    \centering    
    \includegraphics[width=0.9\linewidth]{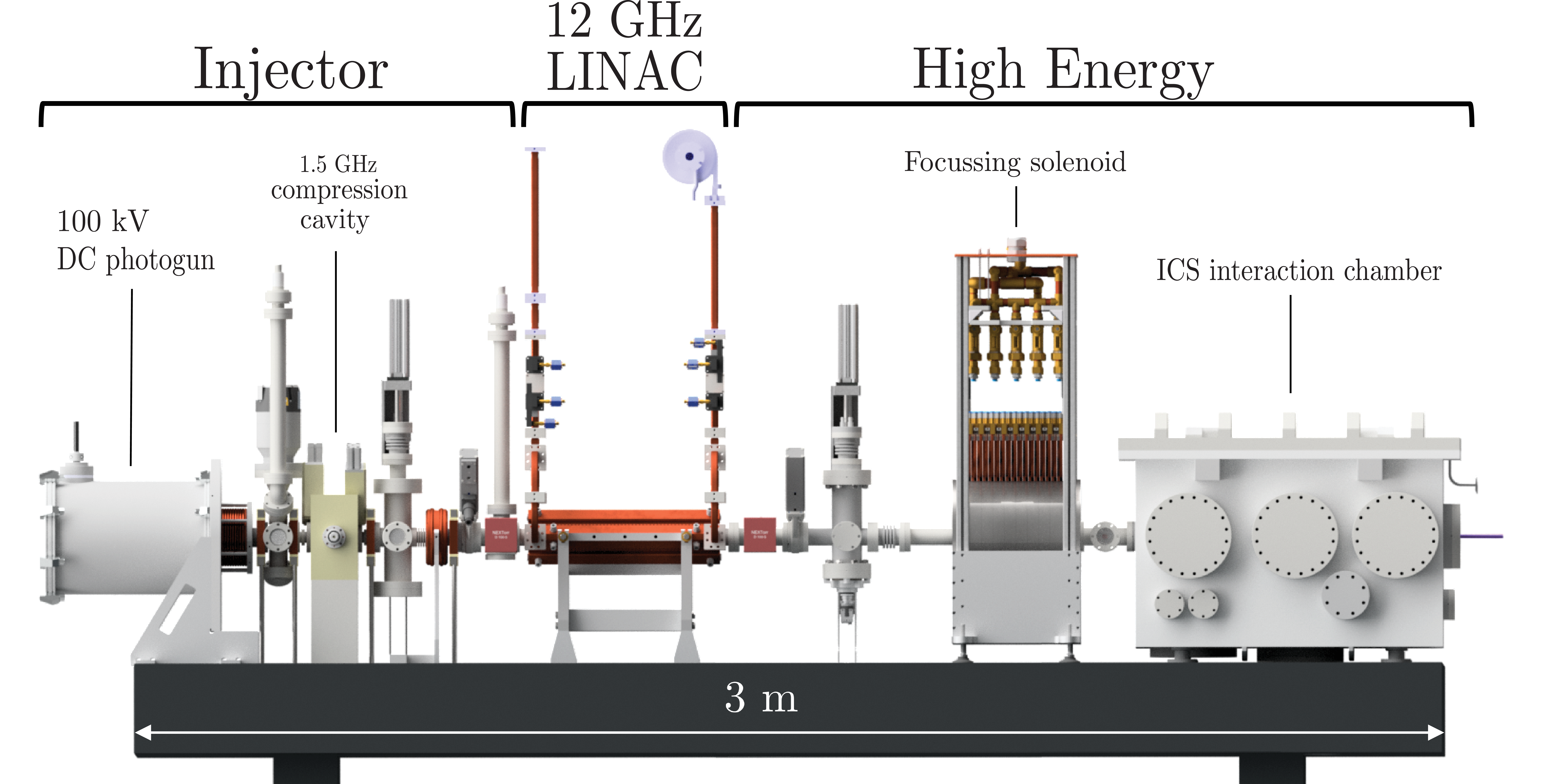}
    \vspace{-0.1cm}
    \caption{The Smart*Light electron beamline with some of the main components indicated.}
    \label{fig:fullbeamline}
\end{figure}

 In this section, each segment of the Smart*Light configuration is discussed separately. A schematic overview of the complete system is shown in Fig. \ref{fig:systemoverview}. The values mentioned on e.g. bunch charge and laser pulse energy are values used to obtain the results. Parameters of which the system should be capable of are discussed in Section \ref{sec:conclusion}. A render of the electron beamline is shown in Fig. \ref{fig:fullbeamline}.

\subsection{Laser}\label{subsec:Laser}

The laser system serves two purposes: generation of electron bunches by photoemission, and the production of x-rays by ICS. To this end, an adapted version of a commercial laser system is used. The seed for the system is the Element$^{\textrm{TM}}$ 2 Ti:Sapph oscillator \cite{element}. The oscillator output is amplified by a Spitfire Ace$^{\textrm{TM}}$ regenerative amplifier, working in combination with two Ascend pump lasers \cite{Spitfire}. The total configuration produces 800 nm pulses with rms pulse lengths of around 100 femtoseconds at 1 kHz repetition rate. Synchronization of the oscillator to the RF phase in the X-band LINAC is done with a Femtolock$^{\textrm{TM}}$ 2 synchronization unit \cite{femtolock}, locking the frequency to an 11.9942 GHz signal supplied by an RF generator. This RF generator acts as the master oscillator for the complete Smart*Light setup. In the Spitfire, part of the 800 nm laser is split and used for the production of UV via third-harmonic generation (THG) for photoemission from a copper cathode. In this manner, the electron bunches and interaction laser pulses are inherently synchronized. Laser pulse energies up to 5 mJ were used for the interaction laser in the experiments described in Section \ref{sec:results}. 

\subsection{Injector segment}

The electron source is a DC photo-emission electron gun (DC photogun), which is used to create 5 pC bunches with an energy of 100 keV by applying field strengths of 10 MV/m to a copper cathode \cite{vanOudheusden_2007,vanOudheusden_2010}. The electrons are extracted from the cathode through photo-emission with 266 nm UV laser pulses, which are generated via THG (see Section \ref{subsec:Laser}). Upon creation of the electron bunches, space charge forces causes a rapid expansion in size, which is counteracted in the transverse direction by a solenoid placed directly behind the photogun. \\
\indent In order to accelerate electron bunches in the X-band LINAC, the 100 keV electron bunch length has to be reduced to the ps level. This is accomplished by a TM$_{010}$ RF compression cavity, acting as a longitudinal lens \cite{Pasmans_2013}. The cavity operates in CW mode at an RF frequency of 1.499275 GHz. The RF phase is actively controlled by a feedback system. The RF signal is the eighth subharmonic of the 11.9942 GHz signal supplied by the master oscillator. Producing electric fields up to 0.5 MV/m, the cavity compresses the electron bunches to approximately 2 ps at the injection point of the accelerator.\\

\subsection{High-power RF segment}\label{subsec:RFsystem}

Electrons are accelerated by a traveling wave LINAC, operating in the $2\pi/3$-mode at a frequency of 11.9942 GHz. The X-band accelerator is designed specifically to allow injection of 100 keV electron bunches. A more detailed design description of the accelerator can be found in \cite{Stragier_2025}. RF pulses supplied to the acceleration structure are generated by a Canon E37123 X-band klystron in combination with a Scandinova K200 modulator. The seed RF signal is derived from the master oscillator. The klystron pulses are compressed and amplified before filling the LINAC by a SLED-I pulse compressor \cite{Woolley_2017}. Ultimately, this combination is capable of producing 500 Hz, 200 ns, 24 MW pulses. In the experiments discussed in this paper, 50 Hz, 200 ns, 6-10 MW average power pulses were used. As a result, accelerating gradients up to 42 MV/m were created, increasing the electron energies from 100 keV up to 21 MeV in less than 0.5 m.\\ 

\subsection{High energy segment and x-ray detection} \label{subsec: HE segment}

\begin{figure}[t]
    \centering    
    \includegraphics[width=0.8\linewidth]{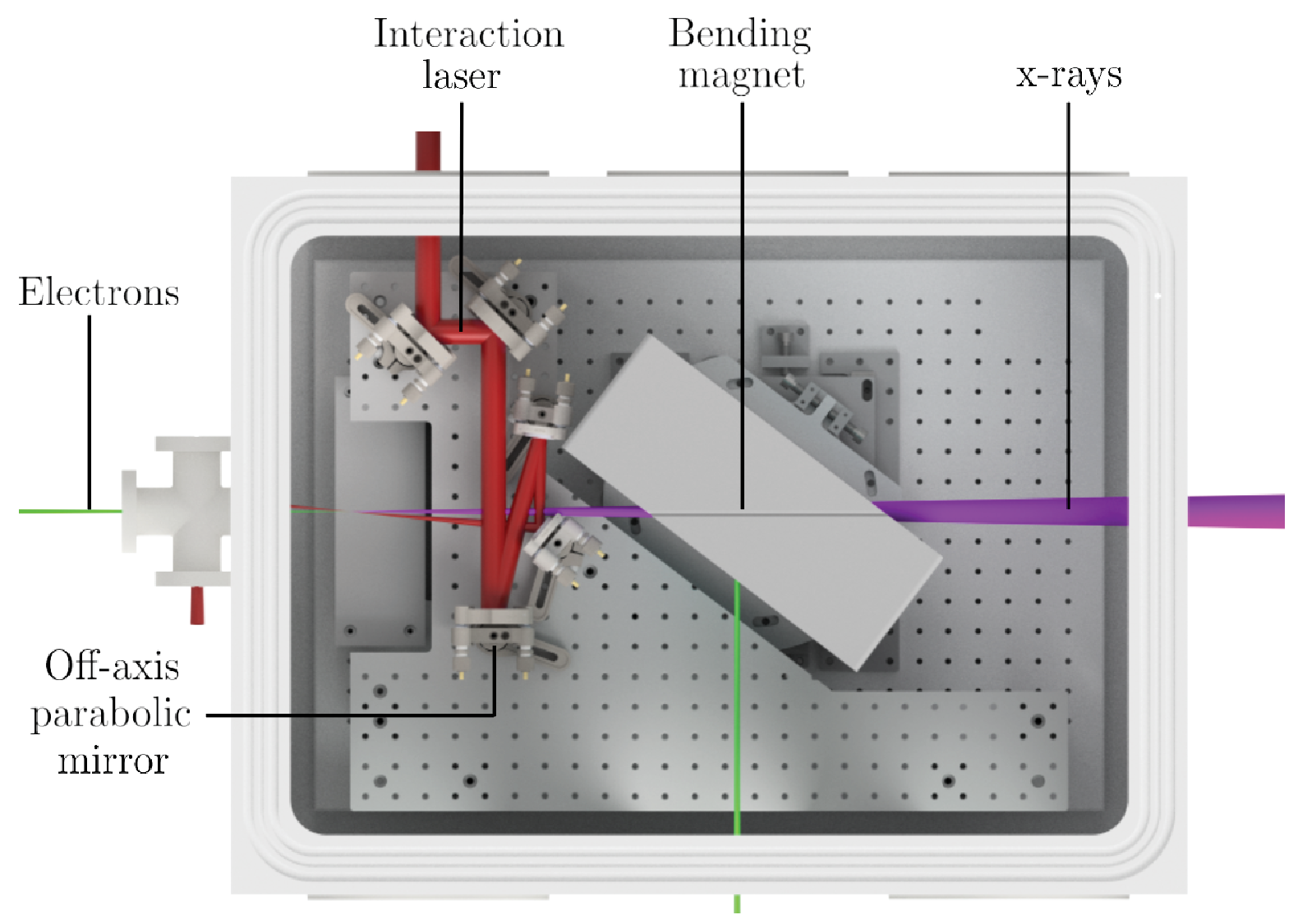}
    \vspace{-0.2cm}
    \caption{Schematic top view of the interaction chamber, showing the path taken by the interaction laser (red) and electrons (green) inside the chamber. The x-rays (purple) pass through the bending magnet, after which they are detected.}
    \label{fig:interactionchamber} 
    \vspace{-0.4cm}
\end{figure}

Before interaction with a laser pulse, the relativistic electron bunch is focused transversely by a solenoid to a small interaction spot in order to increase the x-ray flux, see Eq.\eqref{eq:Nphotons}. After a travel distance of approximately 2.2 m, the bunch arrives at the interaction point (IP), where it collides with an intense femtosecond laser pulse to induce ICS. The laser is focused by an off-axis parabolic mirror. After interaction, electrons are deflected by a permanent dipole magnet into a beam dump. The generated x-ray pulses pass through the gap of the magnet, after which they are detected. For the results, two different detectors have been used. The x-ray intensity distribution has been measured with a MediPix detector \cite{ASI}, while the x-ray energy has been determined using a hyperspectral x-ray camera \cite{Scharf_2011,Boone_2020}. For the measurements with the latter detector a 100 \textmu m thick diamond window was installed in the exit port of the interaction chamber to decouple the vacuum of the chamber and the detector. A top view of the interaction chamber is presented in Fig. \ref{fig:interactionchamber}.

\section{Results}\label{sec:results}

\begin{figure}[ht]
    \centering  
    \vspace{-0.2cm}
    \includegraphics[width=0.95\linewidth]{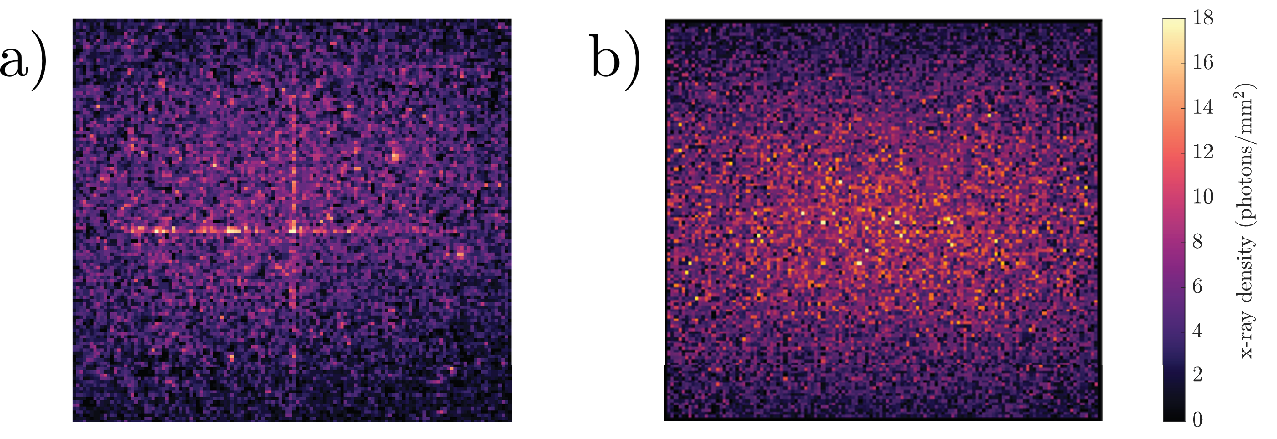}
    \vspace{-0.2cm}
    \caption{a) Typical measurement of the first x-rays results and b) corresponding Monte-Carlo based simulated prediction. Images show an average of 50 shots. Measurements where performed with the 4 quadrant MediPix detector placed 1.2 m from the IP. The complete detection area is 28.2 x 28.2 mm$^2$.}
    \label{fig:firstlight}
\end{figure}

The intensity pattern of the first x-rays generated by the Smart*Light setup is presented in Fig. \ref{fig:firstlight}a. For comparison, the expected signal calculated using a Monte Carlo-based algorithm is depicted in Fig. \ref{fig:firstlight}b \cite{Joos}. The measurement was done using the 4 quadrant MediPix detector. The separation between the 4 quadrants is clearly visible in Fig. \ref{fig:firstlight}a. Although the detector size was too small to completely capture the x-ray beam, the intensity cone can clearly be seen.\\

\indent To compare the photon flux with its theoretical maximum, measurements were done with bunch charges of (5.0 $\pm$ 0.2) pC and laser pulse energies of (4.0 $\pm$ 0.1) mJ. Both electron and laser beams were focused down to $\sigma_e=\sigma_L \approx 20$ \textmu m. The resulting number of x-ray photons per shot was $1.2\cdot 10^3$, while simulations predicted $1.7\cdot 10^3$ photons per shot. The discrepancy can be explained by uncertainties in the experimental parameters.\\
\indent In the following subsections, we elaborate more on the characteristics of the x-ray beam.

\subsection{Short pulse lengths}

\begin{figure}[t]
    \centering   
    \vspace{-0.4cm}
    \includegraphics[width=0.97\linewidth]{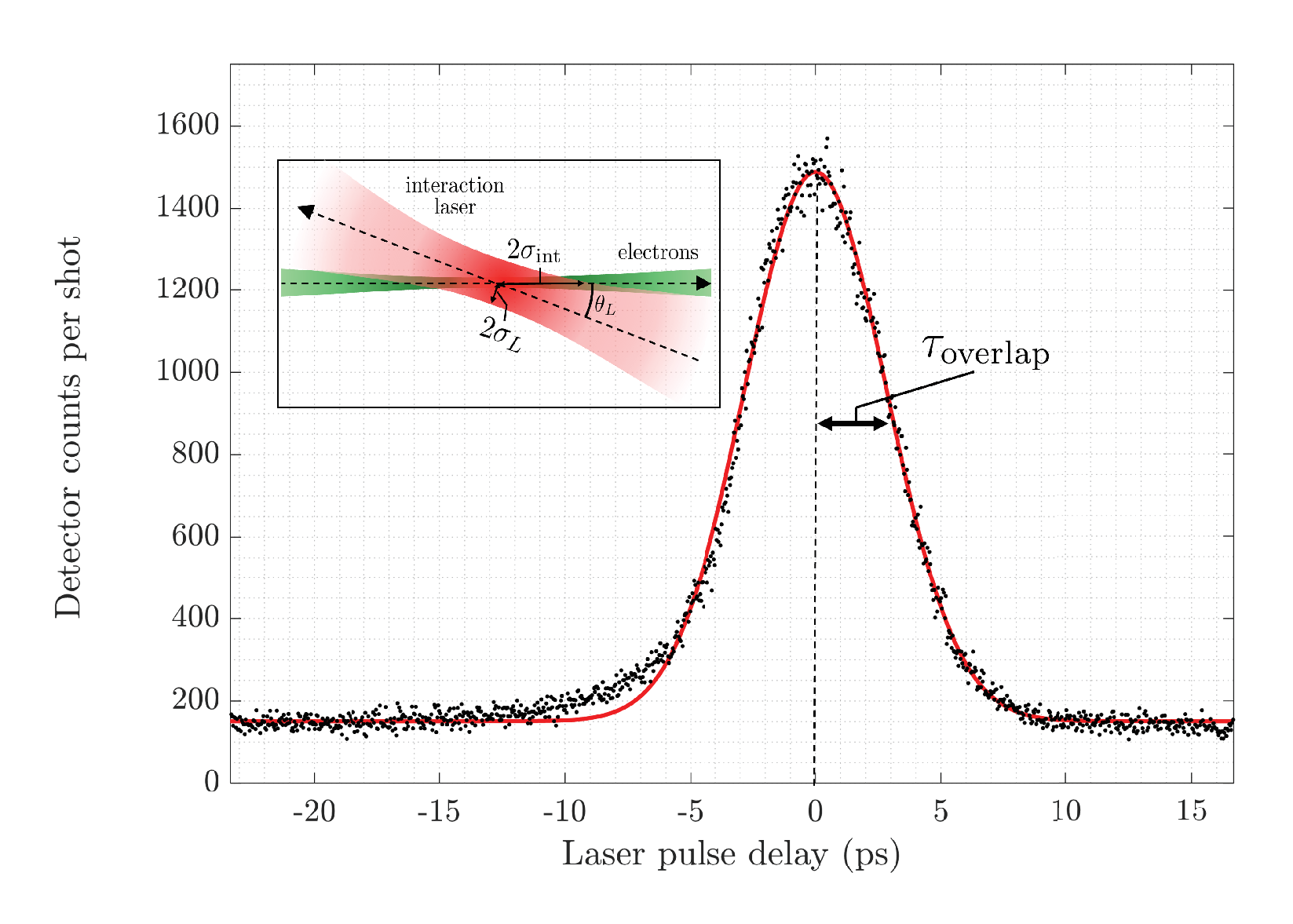}
    \vspace{-0.6cm}
    \caption{Detector counts per shot (black dots) as a function of laser pulse arrival time. The red curve is a Gaussian fit. The inset shows the interaction configuration for $\theta_L = (6 \pm 2)^{\circ}$ applied during the time delay scan.}
    \vspace{-0.3cm}
    \label{fig:delayscan}
\end{figure}

By changing the arrival time of the laser relative to the arrival time of the electron bunches, the x-ray pulse length can be estimated. Both electron and laser beams were focused down to $\sigma_e=\sigma_L \approx 20$ \textmu m, and the laser pulse energy was set to $5.0$ mJ. The Rayleigh length of the laser was $z_R=6.3$ mm. Fig. \ref{fig:delayscan} shows the measured $N_X$ as function of delay time with a Gaussian fit, yielding a rms overlay time $\tau_{\mathrm{overlay}}=2.8$ ps.\\ 
\indent Because the laser pulse length is significantly smaller than the electron pulse length ($\tau_L\ll\tau_e$), $\tau_{\mathrm{overlap}}$ is the result of the convolution between the experienced Gaussian laser profile and electron profile. These measurements were performed with $\theta_L = (6 \pm 2)^{\circ}$, as visualized in the inset in Fig. \ref{fig:delayscan}. The electrons thus experience a Gaussian laser profile with standard deviation $\tau_{\mathrm{int}}= \sigma_\mathrm{int}/c = \sigma_L/(c \sin \theta_L) = (0.7 \pm 0.3)$ ps. The x-ray pulse length $\tau_X$ is equal to $\tau_e$, which is found by $\tau_X=\tau_e=\sqrt{\tau_{\mathrm{overlap}}^2-\tau_{\mathrm{int}}^2}= (2.3 \pm 0.3)$ ps. The uncertainty of the fit was negligible compared to the uncertainty in $\theta_L$. An exact value of $\tau_e$ strongly depends on $\theta_L$ which is experimentally difficult to measure precisely. The results from Fig. \ref{fig:delayscan} do however confirm a $\tau_X$ in the order of picoseconds, with $2.6$ ps as upper bound.

\subsection{Continuous energy tunability}

\begin{figure}[ht]
    \centering   
    \vspace{-0.1cm}
    \includegraphics[width=0.97\linewidth]{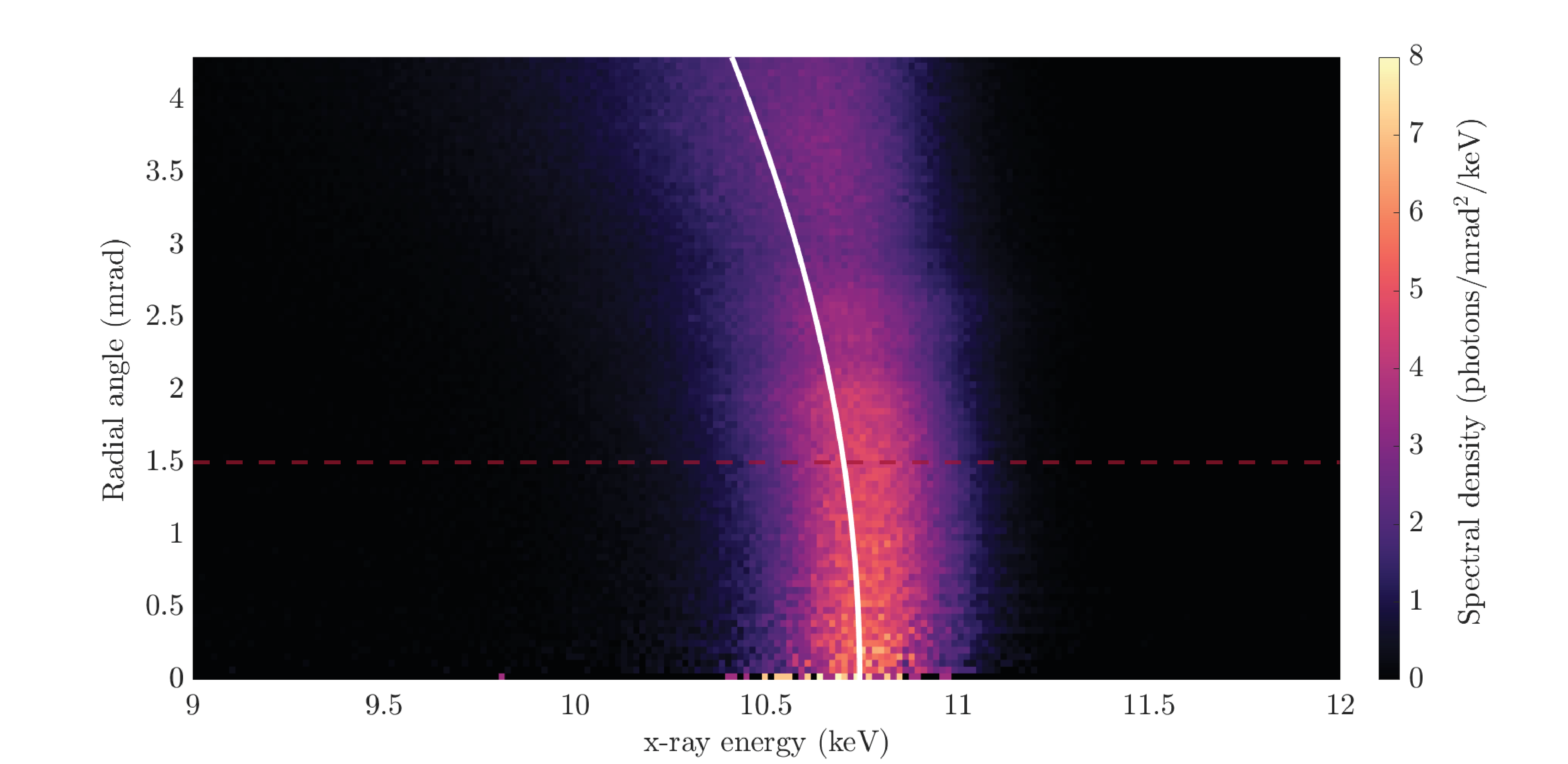}
    \vspace{-0.2cm}
    \caption{The measured spectral angular density distribution for acceleration with 10 MW average power RF pulses supplied to the LINAC. The white line is a reference for the expected shape based on Eq. \eqref{eq:wavelength}.}
    \label{fig:spectraldensity}
    \vspace{-0.1cm}
\end{figure}

To determine the spectral angular density distribution and energy spectrum of the x-ray beam we have used a hyperspectral x-ray camera \cite{Boone_2020}. An example of the measured spectral angular density distribution corresponding to 10 MW average input power supplied to the accelerator is shown in Fig. \ref{fig:spectraldensity}. As reference, Fig. \ref{fig:spectraldensity} shows the expected x-ray energy per radial angle based on Eq. \eqref{eq:wavelength} (white solid curve). For the reference, $\beta$ was found using the x-ray energy corresponding to the peak of the energy spectrum. The behavior  of the measured distribution and Eq. \eqref{eq:wavelength} for different angles $\theta_X$ are comparable. Fig. \ref{fig:spectraldensity} furthermore shows that the x-ray energy spread is uncorrelated with respect to $\theta_X$ for $\theta_x<1.5$ mrad. The following results are therefore based on data acquired within an opening angle $\theta_x=1.5$ mrad.

\newpage

\begin{figure}[htbp]
    \centering    
    \includegraphics[width=1\linewidth]{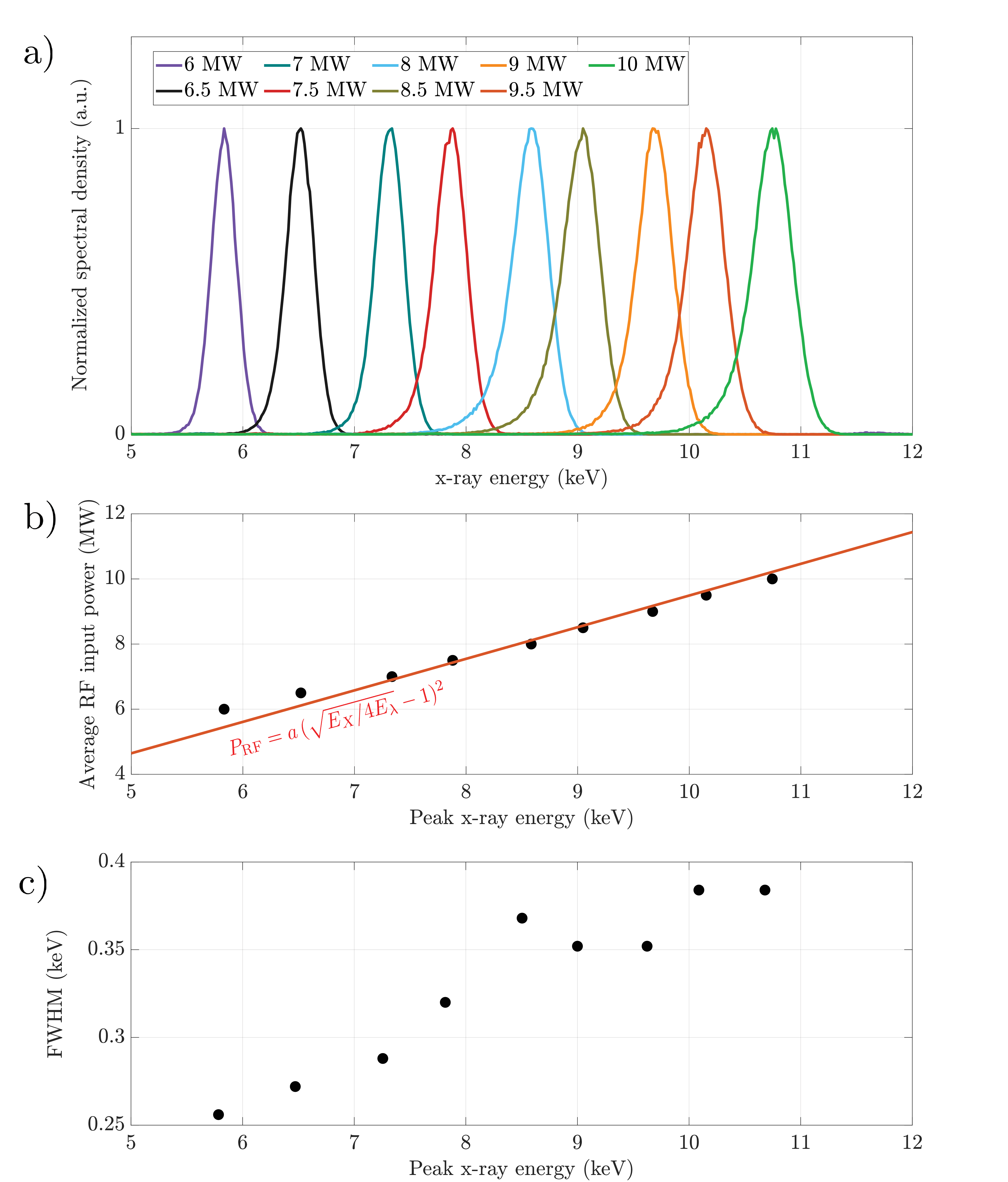}
    \vspace{-0.1cm}
    \caption{Measurements showing the tunability of the source. a) normalized energy spectra for different RF input powers. b) peak x-ray energies of the spectra for the corresponding powers. The red line is a fit with fit parameter $a= 6.2 \cdot 10^{-2}$ MW. c) FHWM per peak x-ray energy. For data acquisition, counts within an opening angle of $\theta_X = 1.5$ mrad were considered.}
    \label{fig:tunability}
\end{figure}

\indent Fig. \ref{fig:tunability}a shows the normalized x-ray energy spectra that has been measured for different average RF input powers supplied to the accelerator, varying the electron energy and thus the x-ray energy. The relation between the average RF input power and the x-ray energy makes it possible to tune the energy in a continuous manner. The spectra before normalization can be found in Supplement 1 (Section S1). The resolution of the used hyperspectral x-ray camera is approximately 144 eV (FWHM at 6 keV) \cite{Boone_2021}.\\
\indent To verify the trend of the generated x-ray spectra as a function of the RF power, the peak x-ray energies of the spectra were plotted for the corresponding average input powers. A fit of the expected trend based on Eq. \eqref{eq:wavelength} is shown alongside the measured data presented in Fig. \ref{fig:tunability}b. The fit parameter corresponded to $a= 6.2 \cdot 10^{-2}$ MW. For the lower peak x-ray energies, the average RF input power is higher than expected. This could be attributed to the fact that acceleration at lower average RF input powers is less efficient. Overall, the trend between x-ray energy and RF power is as expected.\\
\indent Lastly, we have determined the FWHM of the normalized energy spectra presented in Fig. \ref{fig:tunability}a. The results are shown in Fig. \ref{fig:tunability}c, illustrating the monochromaticity of the x-ray beam. The FWHM of the different spectra are in the range of 250 eV - 400 eV, which corresponds to 3.6\% - 4.5\% when normalized to the respective peak energies. The discrepancy of the data point at 8.5 MW compared to the rest could be attributed to instabilities during the measurement. The effect of the opening angle $\theta_X$ on the FWHM for 10 MW average RF input power can be found in Supplement 1 (Section S1). 

\subsection{Polarization}

\begin{figure}[ht]
    \centering    
    \includegraphics[width=1\linewidth]{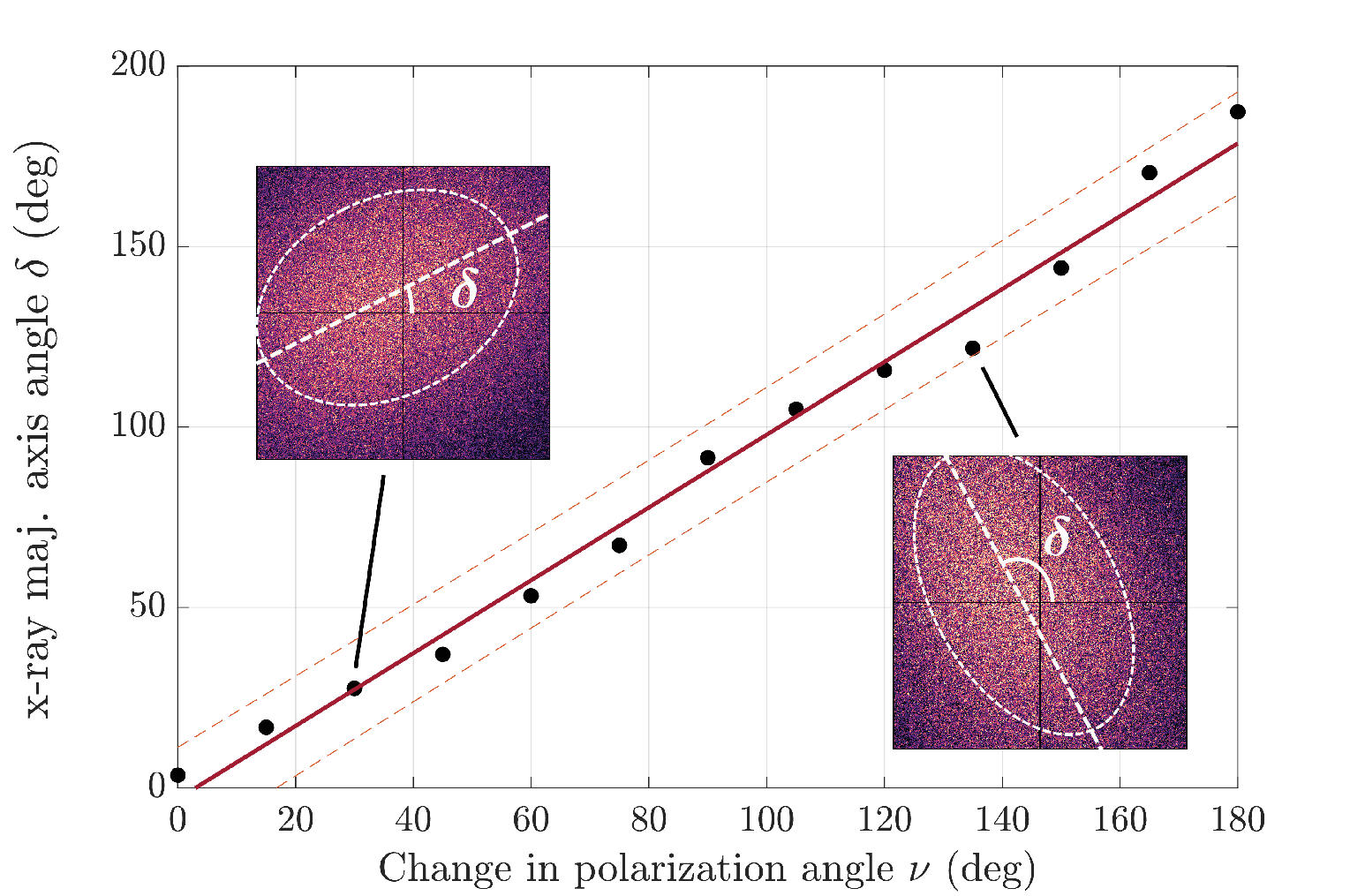}
    \vspace{-0.5cm}
    \caption{Angle between the orientation of the major axis of the spatial x-ray ellipsoid distribution and the horizontal axis as function of the change in incident laser polarization. The red line is a linear fit and the dashed orange lines indicate the 95\% prediction interval. The insets show examples of the spatial x-ray distribution.}
    \label{fig:PCApolscan}
\end{figure}

The polarization of the x-ray photons is directly related to the polarization of the incident laser, which allows for fast polarization control \cite{Brown_2004,Zhang_2024}. To verify this ability, the spatial x-ray distribution was measured for different angles of the incident laser polarization $\nu$. In the case of linear laser polarization, according to Eq. \eqref{eq:polchange} we expect an elliptical-shaped distribution, which rotates with the changing polarization.\\
\indent We have determined the orientation of the major axis and its angle $\delta$ with the horizontal axis of the detector using Principle Component Analysis (PCA). The orientation angle is plotted for the different polarization angles $\nu$ of the incident laser in Fig. \ref{fig:PCApolscan}.  The Fig. shows the linear relation between the two variables, which is as expected. Because it was experimentally difficult to manually specify the orientation of the half-wave plate, the 95\% prediction interval was added to the plot. The insets show examples of the spatial distribution with lines to guide the eye. The results clearly show a rotation in the spatial distribution, confirming a characteristic feature of ICS. Although the polarization of the x-rays is not directly measured, the spatial distributions indirectly indicate the ability to control the polarization. This control is not restricted to linear polarization, as in the measurements, but can also easily be extended to circular polarization.

\section{Conclusion and outlook}\label{sec:conclusion}

In this paper, we have presented the result on the first x-rays generated with the Smart*Light setup. For the experimental parameters during the measurement, the x-ray flux of was $1.2 \cdot 10^3$ photons per shot. Furthermore, we have measured the capability of the Smart*Light setup to generate picosecond pulses, with 2.6 ps as an upper bound. Additionally, we have demonstrated the ability of the setup to tune the x-ray energy in a continuous manner between 5.8 and 10.7 keV. Lastly, it was shown that the polarization of the produced x-rays can be easily controlled. Overall, with the current experimental settings, the results are as expected.\\

\clearpage

\indent Although the setup can already be used for specific experiments, its development is far from complete. By conditioning the HPRF system the repetition rate can be increased from 50 Hz to 500 Hz. Additionally, the system can be conditioned to produce 200 ns, 24 MW pulses, corresponding to an acceleration gradient of 60 MV/m. This will allow for the generation of x-ray energies up to 20 keV. Furthermore, with the current setup the bunch charge and laser pulse energy can be increased to 10 pC and 12 mJ respectively, which will improve the x-ray flux. The flux can be further improved by implementing quadrupoles and an optical telescope, reducing the electron and laser spot sizes. By extrapolating the first results with the future upgrades, the expected average brilliance according to Eq. \eqref{eq:brilliance1} is estimated at  $3\cdot 10^{8}$ photons/(s $\times$ mrad$^2$ $\times$ mm${^{2}}$ $\times$ 0.1\% BW).\\
\indent The modularity of the Smart*Light setup allows for a different electron source. Implementing a thermionic gun operating in burst mode would further increase the expected average brilliance to $10^{12}$ photons/(s $\times$ mrad$^2$ $\times$ mm${^{2}}$ $\times$ 0.1\% BW) \cite{Toonen_2019}. If high coherence is desired over flux, an ultracold electron source would be suitable \cite{Nijhof_2023}. Furthermore, by applying second harmonic generation, the wavelength of the interaction laser can be halved, resulting in a maximum x-ray photon energy of 40 keV. \\
\indent The compactness and relatively low construction and maintenance costs makes the Smart*Light setup ideal as an in-house tunable x-ray source for research institutes, companies, museums, and hospitals. The user can tailor the x-ray output and utilize the continuous energy-tunability and simple polarization control to do their specific experiments, thereby improving on x-ray tubes without the necessity for beam time at a synchrotron or XFEL.

\begin{backmatter}
\bmsection{Funding}
The authors acknowledge financial support from Interreg Vlaanderen-Nederland (Smart*Light project and Smart*Light 2.0 project). This publication is part of the project 'ColdLight: from laser-cooled atoms to coherent soft X-rays' with file number 741.018.303 of the research programme Industrial Partnership Programmes (IPP) which is (partly) financed by the Dutch Research Council (NWO). Co-financed by Holland High Tech (project RT103421 – Hard x-ray source for wafer metrology) 

\bmsection{Acknowledgment}
We like to thank CERN for allowing us to use the drawings of their RF devices and help with fabricating the accelerator. Additionally, thanks to S. Oosterink, H. van den Heuvel and H. van Doorn for their expert technical assistance.

\bmsection{Disclosures}
The authors declare no conflicts of interest.

\bmsection{Supplemental document}
See Supplement 1 for supporting content.

\end{backmatter}

\bibliography{Optica-template}

\end{document}

\maketitle

\section{Continuous tunable energy}

\begin{figure}[ht]
    \centering    
    \includegraphics[width=1\linewidth]{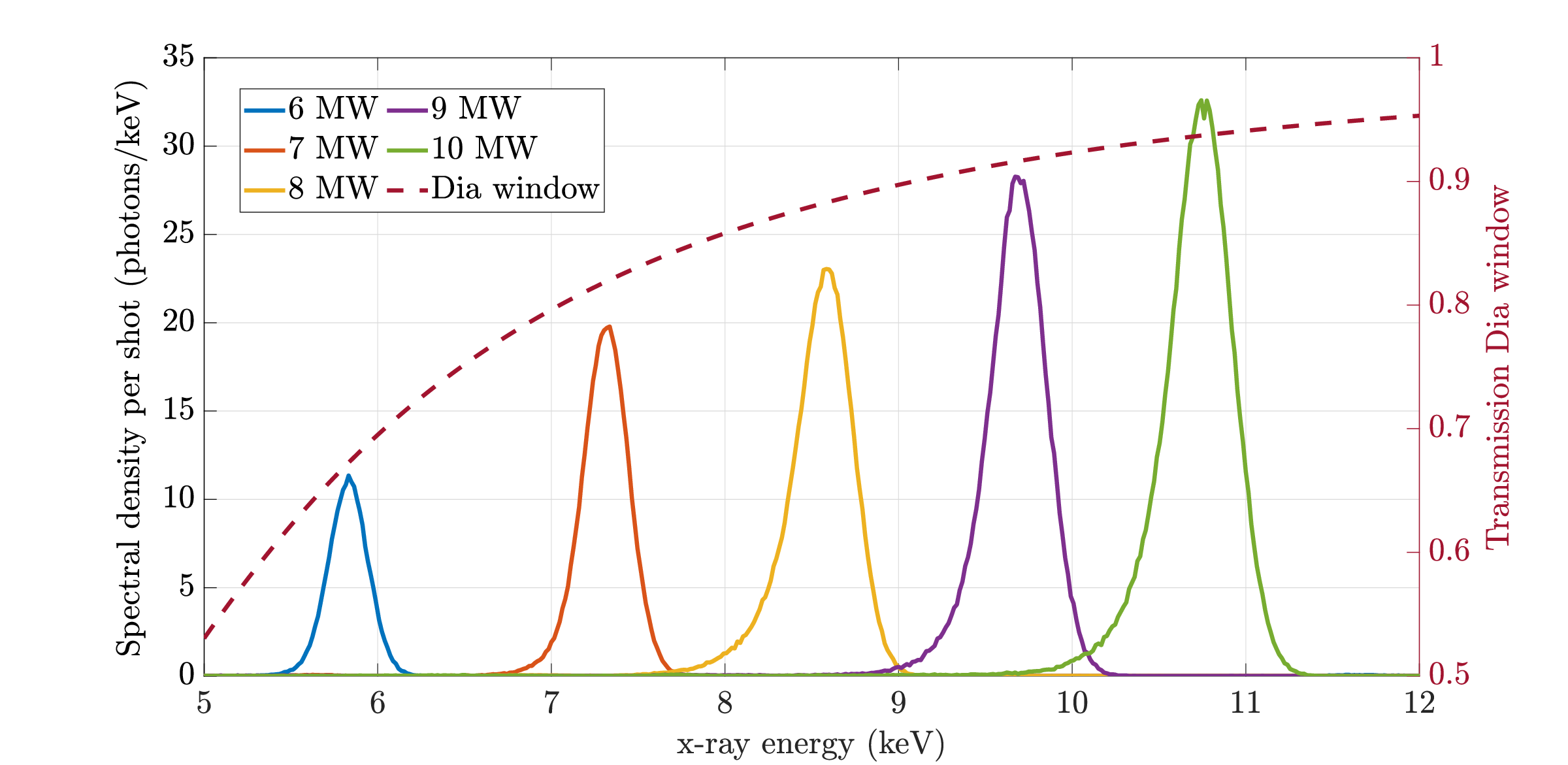}
    \vspace{-0.4cm}
    \caption{Energy spectra measured with a hyperspectral x-ray camera for different average RF input powers. Data was acquired within an opening angle $\theta_X = 1.5$ mrad. Dashed line: Transmission of the diamond window mounted at the back of the interaction chamber.}
    \vspace{-0.6cm}    \label{fig:spectraAppendix}
\end{figure}

\begin{figure}[t]
    \centering    
    \includegraphics[width=1\linewidth]{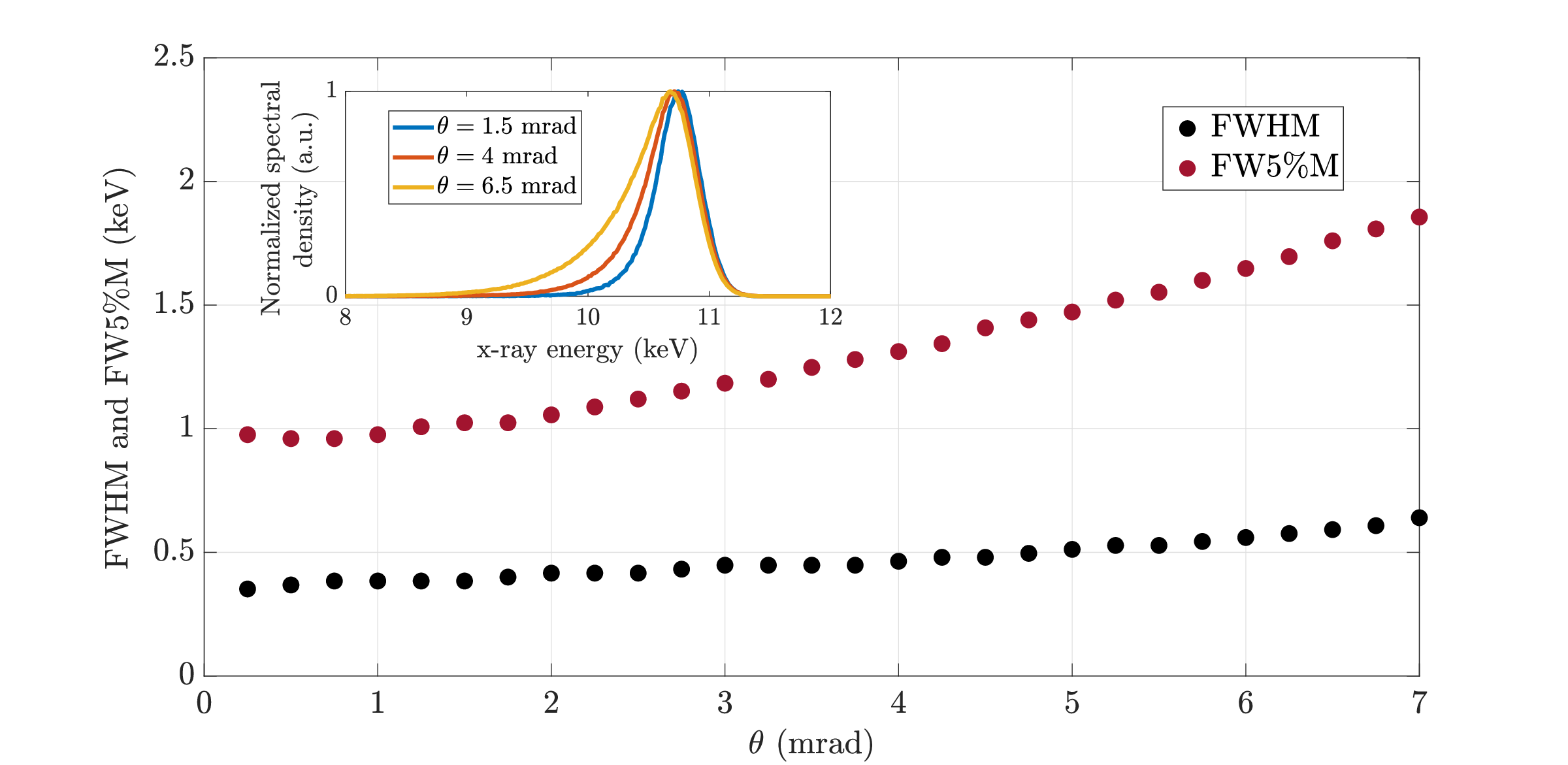}
    \vspace{-0.2cm}
    \caption{FWHM and FW\%5M for different opening angles $\theta$ of the spectra when the accelerator is supplied with an input power of 10 MW. Inset: the spectra for three different opening angles, showing the lower x-ray energy tail. }
    \vspace{-0.6cm}
    \label{fig:FWHMandFWFM}
\end{figure}

The spectra corresponding to Fig. 8 are presented in Fig. \ref{fig:spectraAppendix}. For the acquisition of the spectra, only counts within an opening angle of $\theta = 1.5$ mrad are considered. The energy resolution of the used hyperspectral x-ray camera is approximately 144 eV (FWHM at 6 keV) \cite{Boone_2021}. Clearly, the input power supplied to the accelerator affects the peak x-ray energy, making it possible to choose the latter in a continuous fashion. For lower input powers, the number of measured x-rays reduces. This is mainly a consequence of the use of a 100 \textmu m thick diamond window, which was installed to decouple the vacuum of the interaction chamber and the detector. The transmission of the window has been plotted in Fig. \ref{fig:spectraAppendix}, which corresponds well to the decay in counts for lower x-ray energies.\\
\indent The effect of the chosen opening angle $\theta$ on the FWHM is plotted for the case of 10 MW input power in Fig. \ref{fig:FWHMandFWFM}. In the same figure, the full width at 5\% of the spectra's maxima are plotted (FW5\%M). The latter increases more for larger angles $\theta$ due to the asymmetric lower x-ray energy tail, which is a result of the angular spread of the electron bunch. The growing tail is visualized in the spectra for various $\theta$, added as an inset in Fig. \ref{fig:FWHMandFWFM}.

\vspace{-0.1cm}

%Manual citation list
\bibliography{supplementaldocument}